\documentclass[prl,twocolumn,floatfix,showpacs,epsf,amssymb]{revtex4}
\usepackage{graphicx,subfigure,amsmath,bm,colordvi,times}
 \usepackage[usenames]{color}
 \usepackage{epstopdf}
 \usepackage{multirow}
\begin{document}
 \title{Wigner crystallization in transition metal dichalcogenides: A new approach to correlation energy }
\author{M. Zarenia$^1$, D. Neilson$^{2}$, B. Partoens$^1$, and F. M. Peeters$^1$}
 \affiliation{$^1$Department of Physics, University of Antwerp,
Groenenborgerlaan 171, B-2020 Antwerpen, Belgium\\
$^2$Dipartimento di Fisica, Universit\`a di Camerino, 62032 Camerino, Italy}

\begin{abstract}
We introduce a new approach for the correlation energy of one- and two-valley two-dimensional electron gas (2DEG) systems. Our approach is based on a random phase approximation at high densities and a classical approach at low densities, with interpolation between the two limits. This approach gives excellent agreement with available Quantum Monte Carlo (QMC) calculations. We employ the two-valley 2DEG model to describe the electron correlations in monolayer transition metal dichalcogenides (TMDs).  The zero-temperature transition from a Fermi liquid to a quantum Wigner crystal phase in monolayer TMDs is obtained using density-functional theory within the local-density approximation. Consistent with QMC, we find that electrons crystallize at $r_s=30.5$ in one-valley 2DEG. For two-valleys, we predict Wigner crystallization at $r_s= 29.5$, indicating that valley degeneracy has little effect on the critical $r_s$, in contrast to an earlier claim.
\end{abstract}
\maketitle

%
Research on two-dimensional (2D) materials such as graphene (a 2D allotrope of carbon) and semiconducting transition metal dichalcogenides (TMDs), e.g. MoS$_2$, WS$_2$, MoSe$_2$, WSe$_2$, WTe$_2$, etc., has attracted a lot of interest over the past decade \cite{review,revTMD,rev2}, and is likely to remain one of the leading topics in condensed matter physics and materials science for many years to come. Their interesting properties \cite{review,rev2}, make 2D materials very promising for applications in areas like electronics, sensors, and optical devices. When the effect of electron correlations becomes strong, the properties of the system can undergo profound changes, including novel inhomogeneous ground states \cite{ando,QMC} such as {\it quantum} Wigner crystallization (WC), i.e. freezing of electrons induced by very strong electron-electron interactions \cite{wigner}. Since the experimental evidence of classical WC on the surface of liquid helium in the 1970s \cite{crandall}, the observation of quantum WC in other 2D interacting systems at \textit{zero} magnetic field remains an open experimental challenge.

In a conventional 2D electron gas (2DEG), Quantum Monte Carlo (QMC) calculations predict WC at $r_s\sim31$ \cite{ceperley,dmc-2deg,senatore,needs} which corresponds to an extremely low electron density $\rho\sim 10^{9}$ cm$^{-2}$, requiring very low-disorder samples which to date has not been achievable in solid state systems. The parameter $r_s$, is defined as the ratio of the average electron-electron interaction energy to the Fermi energy.
Graphene is known as a weakly interacting system, since it has a $r_s$ value less than unity independent of the density \cite{sarma}, and is therefore expected not to exhibit any WC \cite{dahal}. However, TMD monolayers with quasi-quadratic low energy dispersions do have a density dependent $r_s\propto1/\sqrt{\rho}$ and this can be large for experimentally accessible densities, e.g. at a charge density of $\rho= 1\times10^{11}$ cm$^{-2}$ in monolayer MoS$_2$, $r_s\approx60$ (see supplementary info of Ref. \onlinecite{TMDrs}). This suggests monolayer TMDs as exciting new candidates for the observation of \emph{quantum} WC.

In 2D monolayer TMDs, inversion symmetry breaking leads to the formation of a direct band gap located at the two inequivalent valleys at the K points (corners of the first Brillouin zone). Therefore the electrons have an additional degree of freedom, i.e. valley index, which can significantly modify the correlation energies. Correlation between the particles interacting with a $1/r$ Coulomb potential, moving in a neutralizing charge background plays a crucial role in determining the WC density. With a lowering of the density, the electrons become strongly correlated and it is the competition between the kinetic energy of the particles and their mutual Coulomb interactions that crystalize the particles. The energy differences between the WC and liquid phases are on a very tiny scale and, to get meaningful predictions, great accuracy such as is afforded by QMC methods is required \cite{dmc-2deg,ceperley,dmc-2v2deg}. Reference \onlinecite{dmc-2v2deg} obtains the correlation energy for a two-valley 2DEG system using QMC calculations to describe the ground state energy of electrons confined in a Si metal-oxide-semiconductor field-effect transistor (MOSFET). They predict that the density of WC shifts to a much lower density, $r_s\approx 45$, making the WC more difficult to form compared to the one-valley system. In the present work we challenge this result.

In our work we first introduce a new approach based on the random phase approximation (RPA) at high densities and a classical approach at low densities, with interpolation between the two limits to obtain the correlation energy in monolayer TMDs, i.e. two-valley 2DEG system. We demonstrate that our results for the correlation energies are in a very good agreement with QMC for both one-valley and two-valley 2DEG systems. Next, we employ a density functional approach and obtain the ground state energy for a non-uniform density distribution corresponding to a 2D hexagonal lattice.
%

There exists an exact expression for the exchange-correlation energy $E_{xc}[\rho]$ \cite{dft,vignale},
\begin{equation}\label{Exc}
E_{xc}[\rho] = \int_{0}^{1}d\alpha~W_{\alpha}[\rho],
\end{equation}
where $W_{\alpha}[\rho]$ is the potential energy functional of $\rho$ excluding  the Hartree contribution, for a fictitious system interacting via a Coulomb-like interaction that is scaled by a multiplicative coupling constant factor $\alpha$.  Since $W_{\alpha}[\rho]$ is expected to be a smooth function of $\alpha$, \cite{Hood} Refs.\ \onlinecite{seidl1,seidl2,pc} proposed an interpolation procedure for $W_{\alpha}[\rho]$, between the $\alpha=0$ (weakly-interacting, high density limit) and $\alpha=\infty$ (strongly-interacting, low density limit),
\begin{equation}\label{eqW0Winf}
  W_{\alpha}[\rho] \simeq W_{\infty} + \frac{W_0-W_{\infty}}{\sqrt{1+2X\alpha}},~~~
  X = \frac{W_0^\prime}{W_\infty-W_0}
\end{equation}
where $W_0^\prime=\text{d}W_\alpha/\text{d}\alpha|_{\alpha=0} $.
Substituting Eq.\ (\ref{eqW0Winf}) into Eq.\ (\ref{Exc}) and integrating,
we obtain for the correlation energy,
\begin{equation}\label{eqExc}
  E_{xc}[\rho] = W_0 + (W_0-W_{\infty})\left[\frac{\sqrt{1+2X}-1}{X}-1\right].
\end{equation}
%
%
\begin{figure}
\hspace{-0.43 cm}
\includegraphics[width=9 cm]{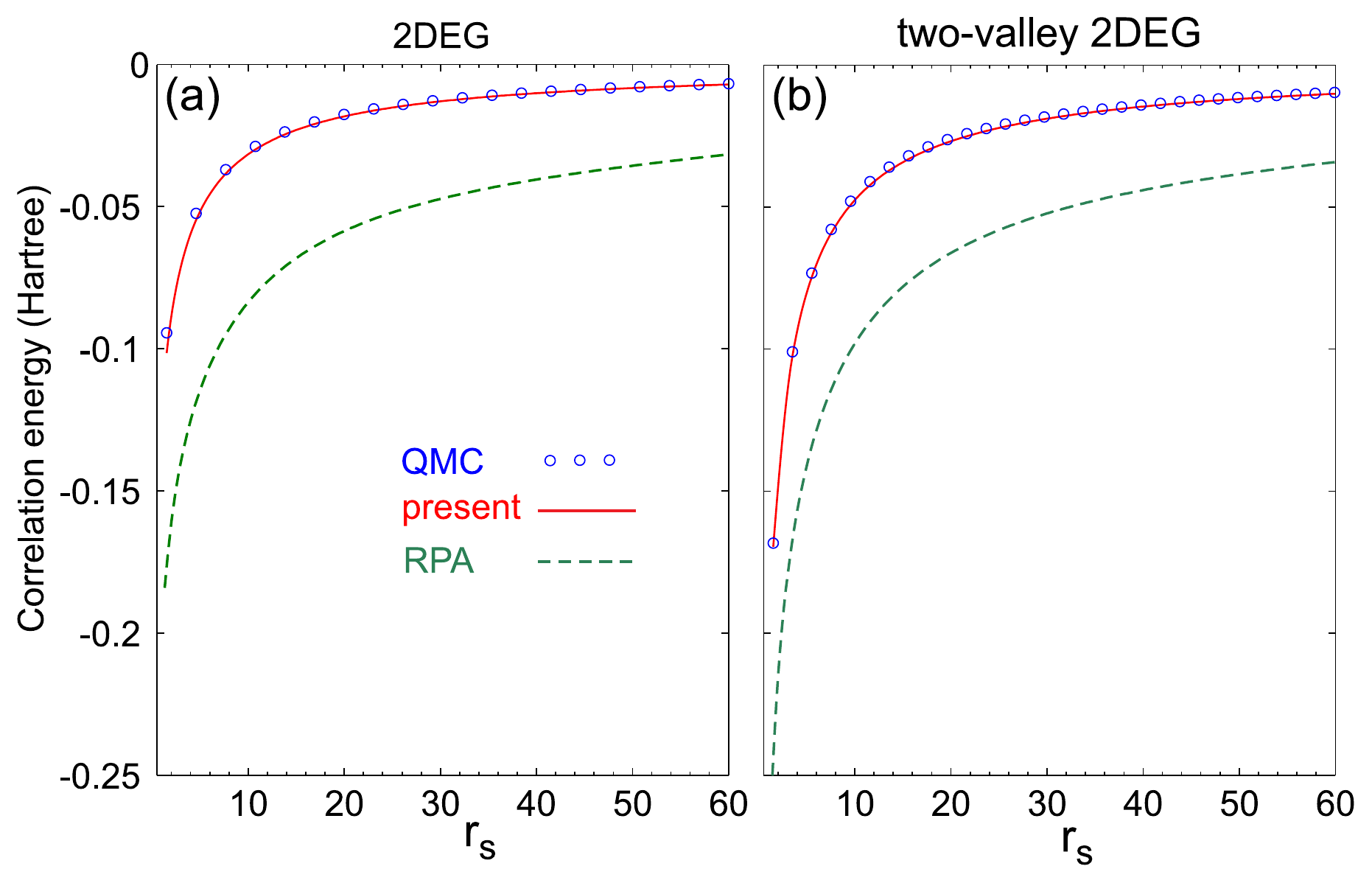}
\caption{Correlation energy of a uniform (a) one-valley 2DEG and (b) two-valley 2DEG systems. The results of the present approach are the solid red curves.  The QMC results (blue open dots) are taken from  Refs.\  \onlinecite{dmc-2deg,ceperley}  for one-valley and Ref.\ \onlinecite{dmc-2v2deg} for two-valley 2DEG systems.  The pure RPA  results are also shown (dashed green curves).}
\label{Fig.1}
\end{figure}

\emph{Weakly-interacting regime.}
In Ref. \onlinecite{seidl2}, $W_0$ is calculated for 2DEG using the Fock integral with the
(occupied) Kohn-Sham orbitals while $W_0'$ is the second order coefficient in G\"{o¨}rling-Levy perturbation theory \cite{levy}.
In our work we instead employ RPA to obtain $W_0$ and $W_0'$.  We recall the RPA is exact in the high-density limit \cite{rpabook}.  Using the RPA offers significant advantages over the approach of Ref.\ \onlinecite{seidl2} since the calculations of $W_0$ and $W_0'$ can be readily generalized to finite temperatures and to non-parabolic energy dispersions.

The exact expressions for $W_0[\rho]$ and $W_0^\prime[\rho]$, are given by the
RPA interaction energy as integrals along the imaginary
frequency axis \cite{rpabook},
\begin{eqnarray}\label{eqW0}
W_0[\rho]\!&=&\!\sum_{\boldsymbol{q}}v_q\left[-\frac{\hbar}{2\pi\rho}
\int_{0}^{\infty} \!\text{d}\omega\chi_{0}(\boldsymbol{q},i\omega)-1\right], \nonumber \\
W_0^\prime[\rho]\!&=&\!-\frac{\hbar}{2\pi\rho}\sum_{\boldsymbol{q}}v_q\int_{0}^{\infty} \!\text{d}\omega
~[\text{d}\chi_\alpha(\boldsymbol{q},i\omega)/\text{d}\alpha]_{\alpha=0},
\end{eqnarray}
where $v_q=e^2/\kappa q$ with $\kappa$ being the dielectric constant, and the RPA $\chi_{\alpha}(\boldsymbol{q},i\omega)$ is,
\begin{eqnarray}\label{eqXl}
\chi_{\alpha}(\boldsymbol{q},i\omega) &=& \frac{\chi_{0}(\boldsymbol{q},i\omega)}{1-\alpha v_q \chi_{0}(\boldsymbol{q},i\omega)}. \nonumber\\
\end{eqnarray}
$\chi_{0}(\boldsymbol{q},i\omega)$ is the dynamic non-interacting density-density response function,
\begin{eqnarray}\label{eqX0}
\chi_{0}(\boldsymbol{q},i\omega) &=& g_sg_v\int d^2\boldsymbol{k}\frac{f_k-f_{\mathbf{k+q}}}{\hbar\omega+\varepsilon_{\boldsymbol{k}}-\varepsilon_{\boldsymbol{k+q}}+i\eta},  \nonumber\\
\end{eqnarray}
where $f_{\boldsymbol{k}}$ is the Fermi function for the wave vector
$\boldsymbol{k}$, $g_s$ and $g_v$ are the spin and valley degeneracies,
$\varepsilon_{\boldsymbol{k}}$ is the lowest energy band, and $\eta>0$ is an infinitesimal number.  For the case of TMD's with $g_s=2$ and $g_v=2$ we assume a parabolic dispersion $\varepsilon_{\boldsymbol{k}}=\hbar^2k^2/2m_e^\ast$, and due to the existence of a large semiconducting gap we neglect the influence of the hole band in the response function.

Using Eqs. (\ref{eqW0}) and (\ref{eqXl}) we obtain
\begin{equation}\label{eqW0eh2}
W_0^\prime[\rho]\! =\! -\frac{\hbar}{2\pi\rho}\sum_{\boldsymbol{q}}v_q^2\int_{0}^{\infty}\!\text{d}\omega \chi_{0}(\boldsymbol{q},i\omega)^2.
\end{equation}
It is crucial to obtain the integrals (\ref{eqW0}) with a high accuracy.  Our computational time needs here are orders of magnitude less than for QMC calculations.

\emph{Strongly-interacting regime.} In the opposite limit of large $\alpha\rightarrow\infty $,  the ground state is the classical WC.
We treat the classical  crystal as a collection of
neutral unit cells, each cell containing its electron, embedded in  a uniform neutralizing background.
To determine the total energy, we represent each unit cell as a disk of diameter $r_0$ with the electron or hole
at its center.  $r_0=1/\sqrt{\pi\rho}$ is the average inter-particle
spacing. Neglecting the interaction between  unit cells, then $W_\infty$ is
the electrostatic energy of a single unit cell consisting of the potential energy between the electron and the positive charged disk $E_{e\oplus}$ plus the self energy of the charged disk $E_{\oplus\oplus}$,
\begin{equation}\label{eqwinfeh}
W_{\infty}=E_{e\oplus}+E_{\oplus\oplus}=-\frac{2}{r_s}+\frac{8}{3\pi r_s}.
\end{equation}
We use Hartree units. Having $W_0$, $W_0^\prime$, and $W_{\infty}$ we obtain the correlation energy using Eq. (\ref{eqExc}).

In  Fig.\  \ref{Fig.1}(a) we compare our results for the correlation energy $E_{c}=E_{xc}[\rho]-W_0$  (red curve) with those obtained from Diffusion QMC (open circles) and pure RPA (green curve) for a one-valley 2DEG system. Comparing with QMC data, adopted from Refs.\ \onlinecite{ceperley} and \onlinecite{dmc-2deg} we see that our approach  exhibits excellent agreement. We also obtained the correlation energy as a function of $r_s$ for a monolayer TMD (Fig.\  \ref{Fig.1}(b)), which we compare  with the corresponding QMC calculations for two-valley 2DEG \cite{dmc-2v2deg}.  We found that the relative errors between the correlation energies of the present approach and QMC are always less than 5\%.

\emph{Ground state energy}. We start by recalling that in Density Functional Theory (DFT),  the ground state energy functional $E[\rho]$ for a system
of interacting electrons with density distribution $\rho(\boldsymbol r)$ is written as the sum,
\begin{equation}\label{eqEt}
E[\rho] = K[\rho]+  E_{\mathrm{coul}}[\rho]+E_{\mathrm{x}}[\rho]+E_\mathrm{c}[\rho],
\end{equation}
where $K[\rho]$, $E_{\mathrm{coul}}[\rho]$, $E_{\mathrm{x}}[\rho]$, and $E_{\mathrm{c}}[\rho]$ respectively denote the non-interacting kinetic-energy, Hartree term,
exchange, and correlation energy functionals.
Within the local-density approximation (LDA),  approximate forms of these functionals in terms of density for
inhomogeneous density distributions are,
\begin{equation}\label{K}
K[\rho]=\frac{1}{8}\int d^2r\frac{\nabla\rho(\boldsymbol{r})\cdot\nabla\rho(\boldsymbol{r})}{\rho(\boldsymbol{r})}+\frac{\pi}{2g_v}\int d^2r \rho(\boldsymbol{r})^2,
\end{equation}
\begin{equation}\label{eqEcoul}
E_{\mathrm{coul}}=\frac{1}{2}\int \text{d}^2r\int  \text{d}^2r'\frac{[\rho(\boldsymbol{r})-\rho_0][\rho(\boldsymbol{r'})-\rho_0]}{|\boldsymbol{r}-\boldsymbol{r'}|},
\end{equation}
\begin{equation}\label{Ex}
E_\mathrm{x}[\rho] =-\frac{4}{3}\sqrt{\frac{2}{\pi g_v}}\int d^2r \rho(\boldsymbol{r})^{3/2},
\end{equation}
\begin{equation}\label{Ec}
E_\mathrm{c}[\rho] =\int d^2r\rho(\boldsymbol{r})\epsilon_c[\rho(\boldsymbol{r}),g_v],
\end{equation}
where $\epsilon_c[\rho_0,g_v]$ is the correlation energy for a system of uniform density $\rho_0$ and $g_v=1,2$ distinguishes the one- and two-valley 2DEG systems. From these expressions, we obtain the total energy per particle, $\epsilon[\rho] =E/\int d^2r \rho_0$, defined in Eq.\ (\ref{eqEt}) for a  uniform density $\rho_0$ and also for
a non-uniform density distribution $\rho(\boldsymbol{r})$ corresponding to a 2D hexagonal lattice. We  determine the total ground state energy $E[\rho]$ from Eqs.\ (\ref{eqEt}), (\ref{K}), and (\ref{eqExc}).

For the WC, we use for the density a variational form of a superposition of
Gaussians centered on the lattice sites,
\begin{figure}
\includegraphics[width=8cm]{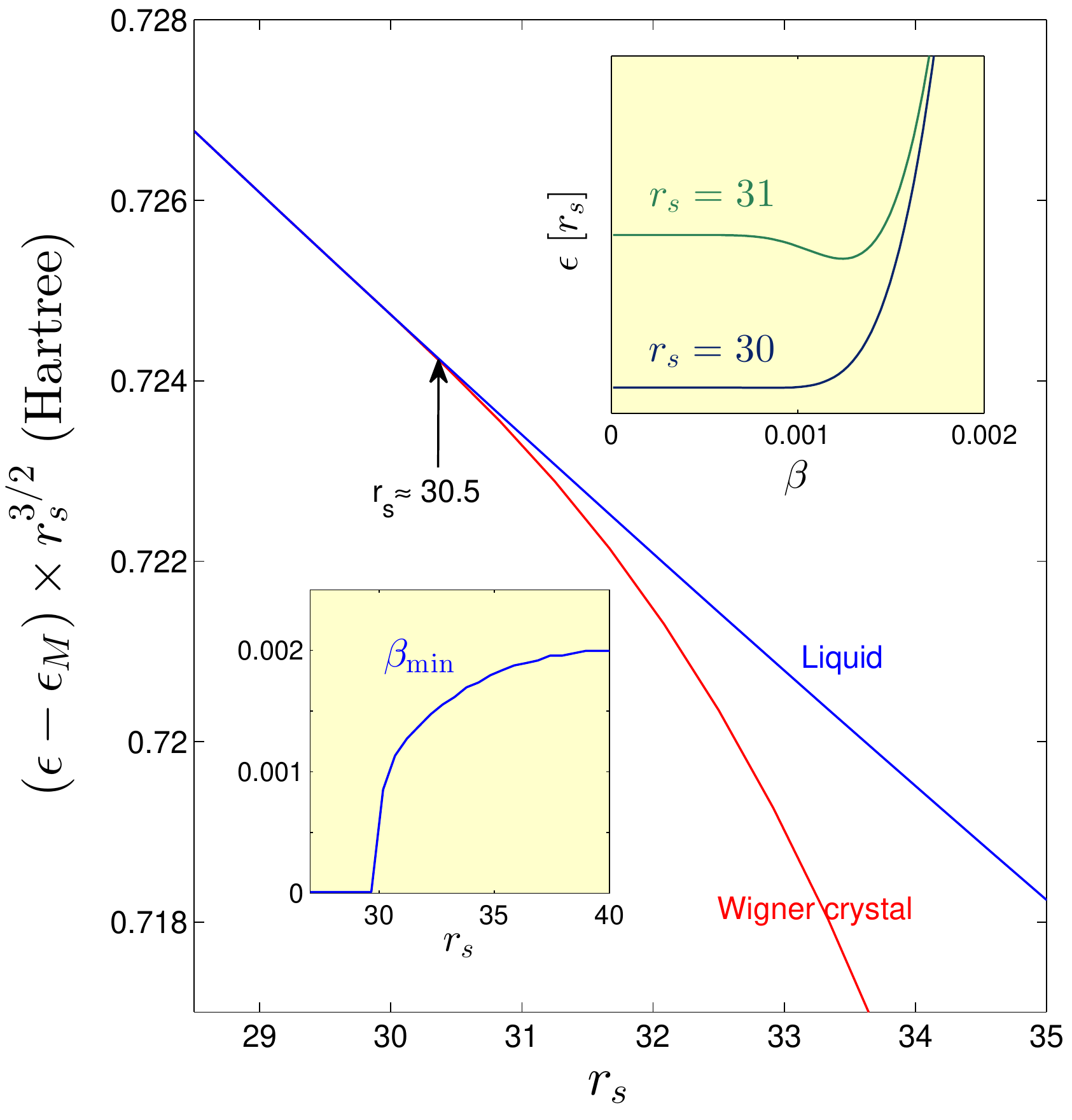}
\caption{The total energy per particle of a one-valley 2DEG system for the liquid (blue curve)
and WC (red curve) phases. The transition to a WC occurs
at $r_s\approx 30.5$, indicated by the vertical arrow in the figure. Upper inset: total energy as a function of variational parameter $\beta$ for $r_s=30$ (blue curve) and $r_s=31$ (green curve). Lower inset: $\beta_{\mathrm{min}}$ as a function of $r_s$. $\epsilon_M=-1.1061/r_s$ is the Madelung energy.}
\label{Fig.2}
\end{figure}
\begin{figure}
\includegraphics[width=8cm]{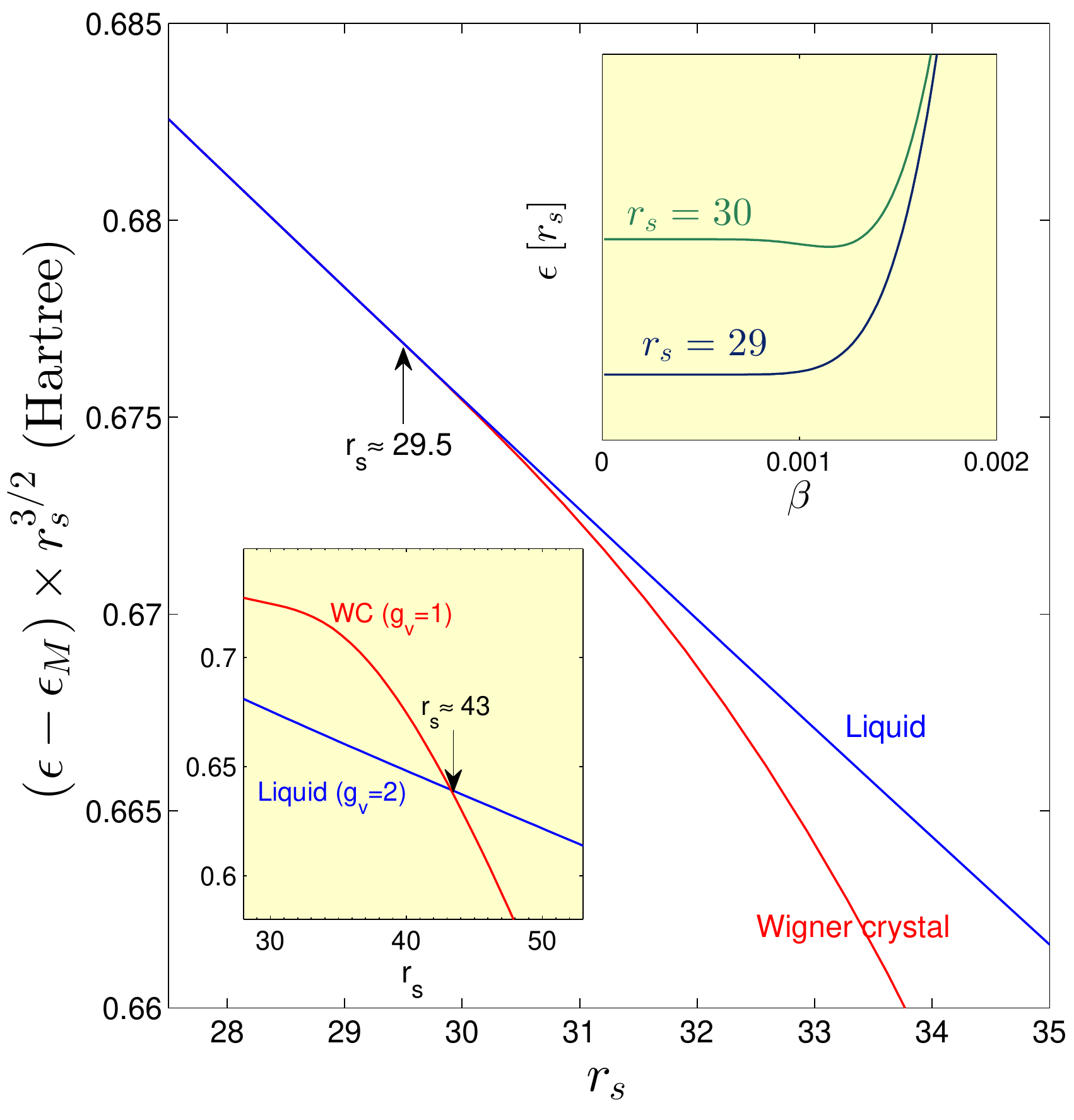}
\caption{The total energy per particle of a two-valley 2DEG system for the liquid (blue curve)
and WC (red curve) phases. The transition to a WC occurs
at $r_s\approx 29.5$, indicated by the vertical arrow in the figure. Upper inset: total energy as a function of variational parameter $\beta$ for $r_s=29$ (blue curve) and $r_s=30$ (green curve). Lower inset: total ground state energies as function of $r_s$ for the $g_v=2$ liquid with a $g_v=1$  WC.}
\label{Fig.3}
\end{figure}
\begin{equation}\label{rhor}
\rho(\boldsymbol{r}) = \frac{\beta}{\pi}\sum_{m,n}\exp[-\beta(\boldsymbol{r}-m\boldsymbol{a}_1-n\boldsymbol{a}_2)^2],
\end{equation}
where $m$ and $n$ are integers and $\boldsymbol{a}_1=(a,0) $ and $\boldsymbol{a}_2=(-a/2,\sqrt{3}a/2) $ are the
lattice vectors for the two-dimensional hexagonal lattice.  $a=\sqrt{2/\sqrt{3}\rho_0}$. The single  variational parameter, $\beta$  determines the degree of localization on the lattice sites. Eq.\  (\ref{rhor}) each Gaussian is isotropic, centered on the lattice points. Equation (\ref{rhor}) may
also be written as a summation of the reciprocal-lattice vectors $\boldsymbol{k}_{mn}$,
viz.
\begin{equation}\label{rhok}
\rho(\boldsymbol{r}) = \rho_0\sum_{m,n}e^{-k_{mn}^2/4\beta}e^{i\boldsymbol{k}_{mn}\boldsymbol{\cdot r}},
\end{equation}
Using this form of density, the Coulomb energy Eq.\  (\ref{eqEcoul}) can be solved analytically. The final
expression of the Coulomb energy per particle is given by,
\begin{equation}\label{eqEcoulk}
\epsilon_{\mathrm{coul}}=\pi\rho_0\sum_{k_{mn}\neq 0}\exp{(-k_{mn}^2/2\beta)}/k_{mn}.
\end{equation}

We numerically calculate the total energy per particle of the ordered phase as follows. For a fixed $r_s$, we calculate
the total energy $\epsilon (\beta)$, with $\beta$ as a variational parameter. We look
for a minimum  in $\epsilon (\beta)$, and use that as the energy
for the inhomogeneous phase. If the minimum is at $\beta=0$, we conclude that the
system is in the liquid state.
The calculated energies for both densities $\rho_0$ and
$\rho(r)$ are plotted as function of $r_s$, in Fig.\ \ref{Fig.2} for a 2DEG system.  We observe a stable WC at $r_s=30.5$,  in excellent agreement with QMC calculations.   (At  $r_s= 30.5$, there is a bifurcation, indicating  a transition to the WC, i.e. the WC energy is lower than the liquid).
The upper inset in Fig. \ref{Fig.2} shows the total energy as a function of the variational parameter $\beta$.   For $r_s=29$ we find  the minimum at $\beta=0$ (liquid phase), but at $r_s=30$, the minimum energy  occurs at a nonzero $\beta\sim 0.0012$ (WC phase).   In the lower inset we show $\beta_{\mathrm{min}}$, the value of $\beta$ at which the energy is minimum, as a function of $r_s$. As $r_s$ increases $\beta_{\mathrm{min}}$ becomes larger leading to  sharply localized peaks on the lattice sites.
\begin{table}
\caption {Wigner crystal electron density of monolayer TMDs at $r_s=30$. The effective electron masses ($m_e^\ast$) are taken from  Ref.\ \onlinecite{mass}, and the out-of-plane (in-plane) dielectric constants $\kappa_\perp$($\kappa_{\parallel}$)  from Ref. \onlinecite{dielectric}.}
\resizebox{0.48\textwidth}{!}{%
\begin{tabular}{c c c c c c c c}
\hline\hline
  TMD & MoS$_2$ & MoSe$_2$ & WS$_2$ & WSe$_2$ & MoTe$_2$ & WTe$_2$\\
 \hline\hline
  $\kappa_\perp$($\kappa_{\parallel}$) & 4.8(3.0) & 6.9(3.8)  & 4.4(2.9) & 4.5(2.9) & 8(4.4) & 5.7(3.3)\\
  $m_{e}^{\ast}$~[$m_{0}$]  & 0.46 & 0.56 & 0.26 & 0.28 & 0.62 & 0.26\\
  $\rho_0$~[$\times10^{11}$ cm$^{-2}$]  & $1.85$ & $1.5$ & $0.66$  & $0.875$ & $1.37$ & $0.45$\\
   \hline\hline
\end{tabular}}
\end{table}

The liquid and WC ground state energies for a two-valley 2DEG system corresponding to monolayer TMDs are shown in Fig.\ \ref{Fig.3}. We observe a transition into WC at $r_s\approx 29.5$ which is in contrast with results obtained using QMC \cite{dmc-2v2deg}.  Reference \onlinecite{dmc-2v2deg} finds the WC transition at $r_s\approx 45$ which corresponds to a much lower transition density than  for the one-valley 2DEG.  However in Ref. \onlinecite{dmc-2v2deg} the authors assume that the WC energy is independent of the number of additional components (spin and valley), and since the liquid energy reduces with an additional valley (because the kinetic, exchange, and correlation energies reduce), it crosses the WC energy at a larger $r_s$. In the lower inset of Fig. \ref{Fig.3} we demonstrate that the $g_v=1$ WC ground state energy does indeed cross the two-valley  ($g_v=2$) liquid energy at $r_s\approx43$, consistent with  the prediction in Ref. \onlinecite{dmc-2v2deg}. But the WC energy is in fact not independent of the number of valleys. This is confirmed by comparing QMC results for the unpolarized and fully-polarized 2DEG in the WC regime (compare the WC energies at $\xi=0$ (unpolarized) and $\xi=1$ (fully-polarized) in Fig.\ 5 of Ref.\ \onlinecite{ceperley}). Using the analogy of these significant WC energy differences between one and two spin components, we expect the same energy difference for the WC in one- and two-valley systems.

The results for the corresponding electron density at $r_s=30$, i.e.\ in the WC regime, are shown in Table I for different TMDs. The electron density is calculated using $\rho_0=1/\pi(a_B^\ast r_s)^2$ where $a_{B}^{\ast}=\hbar^2\kappa/e^2m^{\ast}_e$ with $\kappa=\sqrt{\kappa_{\perp}\kappa_{\parallel}}$.  We find that $r_s=30$ corresponds to the carrier densities $\rho_0 \gtrsim 0.4\times10^{11}$ cm$^{-2}$ for the TMD monolayers of Table I, which is within the lower limit of experimentally attainable densities for monolayer TMDs (of the order of $\rho \agt 10^{10}$ cm$^{-2}$, e.g. see Refs. \onlinecite{exp1,exp2,exp3}).

In summary, we have presented a new approach to obtain the correlation energy in 2D systems. The approach is based on  an interpolation of the exchange-correlation energy between the weakly-interacting  regime and the strongly-interacting classical regime. For the weakly-interacting limit we employed RPA which is exact in this limit.  For the WC regime at low densities, with the electrons forming a lattice, we obtained the electrostatic energy for an assembly of unit-cells that are neutralized by a uniformly charged background. We calculated the correlation energies for one- and two-valley 2DEG systems. The present approach is extremely fast and provides accurate results in excellent agreement with QMC. The approach can readily be extended to: \emph{i}) 2D systems with non-parabolic low energy bands such as one present in few-layer graphene and \emph{ii}) to non-zero temperatures.

We utilized density-functional theory within the local-density approximation to investigate the zero-temperature transition from a Fermi liquid to WC phase.
We find a stable WC at $r_s\agt 30.5$ for the one-valley 2DEG and at $r_s\agt 29.5$ for the two-valley 2DEG.
In monolayer TMDs, $r_s= 29.5$ corresponds to density $\rho_0\gtrsim 0.4\times10^{11}$ cm$^{-2}$, which lies well within the experimentally achievable density range.
We conclude that monolayer TMDs could be ideal systems for the observation of WC in zero magnetic field.

{\it {\bf Acknowledgements.}} This work was partially supported by the Flemish Science
Foundation (FWO-Vl). D.N. acknowledge support by the University of Camerino
FAR project CESEMN.


\begin{thebibliography}{99}


\bibitem{review}
A. H. Castro Neto, F. Guinea, N. M. R. Press, K. S. Novoselov, A. K. Geim, Rev. Mod. Phys. {\bf 81}, 109 (2009).

\bibitem{revTMD}
M. Chhowalla, Zh. Liu, and H. Zhang, Chem. Soc. Rev. {\bf 44}, 2584-2586 (2015).

\bibitem{rev2}
Z. Sheneve {\it et al.}, ACS Nano {\bf 7}, 2898–2926 (2013).

\bibitem{ando} T. Ando, A.B. Fowler and F. Stern, Rev. Mod. Phys. {\bf 54}, 437 (1982).

\bibitem{QMC}
W. M. C. Foulkes, L. Mitas, R. J. Needs, and G. Rajagopal, Rev. Mod. Phys. {\bf 73}, 33 (2001).

\bibitem{wigner}
E. Wigner, Phys. Rev. {\bf 46}, 1002 (1934).

\bibitem{crandall}
R. S. Crandall and R. Williams, Phys. Lett. A {\bf 34}, 404 (1971).

\bibitem{dmc-2deg}
C. Attaccalite, S. Moroni, P. Gori-Giorgi, and G. B. Bachelet, Phys. Rev. Lett. {\bf 88}, 256601 (2002);

\bibitem{ceperley}
B. Tanatar and D. M. Ceperley, Phys. Rev. B {\bf 39}, 5005 (1989).

\bibitem{senatore}
F. Rapisarda, G. Senatore, Australian Journal of Phys. {\bf 49}, 161 (1996).

\bibitem{needs}
N. D. Drummond, R. J. Needs, Phys. Rev. Lett. {\bf 102}, 126402 (2009).

\bibitem{sarma}
S. Das Sarma, Sh. Adam, E. H. Hwang, and Enrico Rossi, Rev. Mod. Phys. {\bf 83}, 407 (2011).

\bibitem{dahal}
H. P. Dahal, Y. N. Joglekar, K. S. Bedell, and A. V. Balatsky, Phys. Rev. B {\bf 74}, 233405 (2006).

\bibitem{TMDrs}
K. F. Mak, K. He,	Ch. Lee,	G. H. Lee,	J. Hone,	T. F. Heinz, and J. Shan, Nature Materials {\bf 12}, 207–211 (2013).

\bibitem{dmc-2v2deg}
M. Marchi, S. De Palo, S. Moroni, and G. Senatore, Phys. Rev. B {\bf 80}, 035103 (2009).

\bibitem{dft}
R. G. Parr and W. Yang, Density-Functional Theory of Atoms
and Molecules, Oxford University Press, New York (1989); R.
M. Dreizler and E. K. U. Gross, Density Functional Theory
Springer, Berlin, (1990).

\bibitem{vignale}
P. Gori-Giorgi, M. Seidl, and G. Vignale, Phys. Rev. Lett. {\bf 103}, 166402 (2009).

\bibitem{Hood} R. Q. Hood, M. Y. Chou, A. J. Williamson, G. Rajagopal, R. J. Needs, and W. M. C. Foulkes,
Phys.\ Rev.\ Lett.\ {\bf 78}, 3350 (1997);
M. Levy and J. P. Perdew, Phys.\ Rev.\ A {\bf 32}, 2010 (1985).

\bibitem{seidl1}
M. Seidl, Phys. Rev. A {\bf 60}, 4387 (1999).

\bibitem{seidl2}
M. Seidl, J. P. Perdew, and M. Levy, Phys. Rev. A {\bf 59}, 51 (1999).

\bibitem{pc}
M. Seidl and J. P. Perdew, Phys. Rev. B {\bf 50}, 5744 (1994).

\bibitem{rpabook}
G. Giuliani and G. Vignale, Quantum Theory of the Electron Liquid, Cambridge University Press, New York (2008).

\bibitem{kimball}
A. K. Rajagopal and John C. Kimball, Phys. Rev. B {\bf 15}, 2819 (1977).

\bibitem{levy}
A. G\"{o¨}rling and M. Levy, Phys. Rev. B {\bf 47}, 13105 (1993).

\bibitem{mass}
A. Korm\'{a}nyos, G. Burkard, M. Gmitra, J. Fabian, V. Z\'{o}lyomi, N. D. Drummond, and V. Fal\'{}ko, 2D Mater. {\bf 2}, 049501 (2015).

\bibitem{dielectric}
A. Kumar and P. K. Ahluwalia, Physica B: Condensed Matter {\bf 407}, 4627 (2012).

\bibitem{exp1}
K. F. Mak, K. He, J. Shan, and T. F. Heinz, Nat. Nanotechnol. {\bf 7}, 494 (2012); J. S.
Ross ,S. Wu, H. Yu,	N. J. Ghimire,	A. M. Jones, G. Aivazian, J. Yan, D. G. Mandrus, Di Xiao, W. Yao, and X. Xu, Nature Commun. {\bf 4}, 1474 (2013).

\bibitem{exp2}
K. F. Mak, K. L. McGill, J. Park, and P. L. McEuen Science  {\bf 344}, 1489 (2014).

\bibitem{exp3}
S. B. Desai, S. R. Madhvapathy, A. B. Sachid, J. P. Llinas, Q. Wang, G. H. Ahn, G. Pitner, M. J. Kim, J. Bokor, Ch. Hu, H. -S. Ph. Wong, A. Javey, Science {\bf 354}, 99 (2016).



\end{thebibliography}
\end{document}